\documentclass[12pt]{iopart}
\expandafter\let\csname equation*\endcsname\relax
\expandafter\let\csname endequation*\endcsname\relax
\usepackage{amsfonts} 
\usepackage{amsmath}
\usepackage{iopams}
\usepackage{amssymb}
\usepackage{mathtools}
\usepackage{cancel} 
\usepackage{bm}
\usepackage[usenames, dvipsnames]{color} 
\usepackage{dcolumn}
\usepackage{epic} 
\usepackage{epsfig}
\usepackage{feynmp}
\usepackage{graphicx}
\usepackage{grffile}
\usepackage[breaklinks,colorlinks = true,linkcolor = red,urlcolor=blue,citecolor=red]{hyperref}
\usepackage{mathrsfs}
\usepackage{soul}
\usepackage{wrapfig}
\usepackage{xy} 
\usepackage{xcolor}
\usepackage{amsmath,amssymb,amsfonts,amsthm}
\usepackage[ascii]{inputenc}
\usepackage{makecell}
\allowdisplaybreaks

\newcommand{\be}{\begin{equation}}
	\newcommand{\ee}{\end{equation}}
\newcommand{\ba}{\begin{aligned}}
	\newcommand{\ea}{\end{aligned}}
\newcommand{\bea}{\begin{eqnarray}}
	\newcommand{\eea}{\end{eqnarray}}	
	
\newcommand{\beal}{\begin{align}}
\newcommand{\eal}{\end{align}}

\newcommand{\blue}{\textcolor{black}}

\newcommand{\bx}{{\bf x}} 

\graphicspath{{images/}{../Figures}}

\begin{document}
%%%%%%%%%%%%%%%%%%%%%%%%%%%

%%%%%% TITLE %%%%%%%%%%%%%%
\title{Linear statistics for Coulomb gases: higher order cumulants}
%%%%%%%%%%%%%%%%%%%%%%%%%%%

%%%%%% AUTHORS %%%%%%%%%%%%
%\author{Jitendra Kethepalli, Manas Kulkarni, Anupam Kundu, Satya N. Majumdar, David Mukamel and Gregory Schehr}
%\institute{
%Jitendra Kethepalli \at International Centre for Theoretical Sciences, Tata Institute of Fundamental Research, Bengaluru -- 560089, India
%\and
%Manas Kulkarni \at International Centre for Theoretical Sciences, Tata Institute of Fundamental Research, Bengaluru -- 560089, India
%\and
%Anupam Kundu \at International Centre for Theoretical Sciences, Tata Institute of Fundamental Research, Bengaluru -- 560089, India
%\and
%Satya N. Majumdar \at LPTMS, CNRS, Univ.  Paris-Sud,  Universite Paris-Saclay,  91405 Orsay,  France
%\and
%David Mukamel \at Department of Physics of Complex Systems, Weizmann Institute of Science, Rehovot 7610001, Israel
%\and
%Gregory Schehr \at LPTMS, CNRS, Univ.  Paris-Sud,  Universite Paris-Saclay,  91405 Orsay,  France \\ Sorbonne Universite, Laboratoire de Physique Theorique et Hautes Energies, CNRS UMR 7589, 4 Place Jussieu, 75252 Paris Cedex 05, France
%}
\author{Benjamin De Bruyne}
\address{LPTMS, CNRS, Univ.  Paris-Sud,  Universit\'e Paris-Saclay,  91405 Orsay,  France}
%\author{Abhishek Dhar}
%\address{International Centre for Theoretical Sciences, Tata Institute of Fundamental Research, Bengaluru -- 560089, India}
%\author{Manas Kulkarni}
%\address{International Centre for Theoretical Sciences, Tata Institute of Fundamental Research, Bengaluru -- 560089, India}
%\author{Anupam Kundu}
%\address{International Centre for Theoretical Sciences, Tata Institute of Fundamental Research, Bengaluru -- 560089, India}
\author{Pierre Le Doussal}
\address{Laboratoire de Physique de l'Ecole Normale Sup\'erieure, CNRS, ENS and PSL Universit\'e, Sorbonne Universit\'e, Universit\'e Paris Cit\'e,
24 rue Lhomond, 75005 Paris, France}
\author{Satya N. Majumdar}
\address{LPTMS, CNRS, Univ.  Paris-Sud,  Universit\'e Paris-Saclay,  91405 Orsay,  France}
\author{Gr\'egory Schehr}
\address{Sorbonne Universit\'e, Laboratoire de Physique Th\'eorique et Hautes Energies, CNRS UMR 7589, 4 Place Jussieu, 75252 Paris Cedex 05, France}

%%%%%%%%%%%%%%%%%%%%%%%%%%%

%%%%%%%% ABSTRACT %%%%%%%%%

\begin{abstract}
We consider $N$ classical particles interacting via the Coulomb potential in spatial dimension $d$
and in the presence of an external trap, at equilibrium at inverse temperature $\beta$. 
In the large $N$ limit, the particles are confined within a droplet of finite size. We study smooth  
linear statistics, i.e. the fluctuations of sums of the form ${\cal L}_N = \sum_{i=1}^N f(\bx_i)$, where $\bx_i$'s
are the positions of the particles and where $f(\bx_i)$ is a sufficiently regular function. 
There exists at present standard results for the first and second moments of ${\cal L}_N$ in the large $N$ limit, as well as associated
Central Limit Theorems in general dimension and for a wide class of confining potentials. 
Here we obtain explicit expressions for the higher order cumulants of ${\cal L}_N$ at large $N$,
when the function 
$f(\bx)=f(|{\bf x}|)$ and the confining potential are both rotationnally invariant. 
A remarkable feature of our results is that these higher cumulants depend only on the value of $f'(|\bf x|)$ and its higher
order derivatives {\it evaluated exactly at the boundary of the droplet}, which in this case is a $d$-dimensional sphere. 
In the particular two-dimensional case $d=2$ at the special value $\beta=2$, a connection to the Ginibre ensemble allows us to
derive these results in an alternative way using the tools of determinantal 
point processes. Finally we also obtain the large deviation form of the full probability
distribution function of ${\cal L}_N$. 
\end{abstract}
%\maketitle
\date{\today}
%%%%%%%%%%%%%%%%%%%%%%%%%%%
\maketitle
%\tableofcontents
%%%%%%%%%%%%%%%%%%%%%%%%%%%

\maketitle

\section{Introduction and main results}\label{Introduction}

\subsection{Background}

During the last decades, there has been a growing interest in the study of classical long-range interacting systems, both in statistical 
physics \cite{Dauxois} and in mathematics \cite{lewin}. An emblematic model of such systems is the so-called ``Coulomb gas'', where $N$ particles interact via the pairwise Coulomb potential in $d$ dimensions [see Eq. \eqref{Coulomb_int} below] and in the presence of an external trap. The thermodynamic properties of these systems have been studied a long time ago in the physics literature, in particular in the context of plasma physics
\cite{Lenard,Baxter1963,AM1980,Choquard1981,JLM93,Janco95,forrester_el}. More recently, there has been a vivid renewed interest for Coulomb gases, motivated in particular by their connections, in $d=2$, to non-Hermitian random matrices 
\cite{Mehtabook,Forrester},
as well as to the quantum Hall physics \cite{Cooper,Charles,oblak}, in particular in the context of non-interacting fermions in a rotating trap
\cite{Lacroix_rotating,Smith_rotating,Manas_rotating1,Manas_rotating2}. \blue{An important class of 
observables that have generated a lot of interest is called {\it linear statistics}, denoted by ${\cal L}_N$ and defined as ${\cal L}_N = \sum_{i=1}^N f({\bx_i})$ where the $\bx_i$'s denote the positions of the particles and where $f(\bx)$ is an arbitrary function. For instance, for $f(\bx) = \bx$, the linear statistics ${\cal L}_N$ simply denotes the position of the center of mass of the gas, but more general functions can be considered. For instance if $f(\bf x)$ denotes the indicator function of some region $\Omega$ of space, then ${\cal L}_N$ denotes the number of particles inside ${\Omega}$, which is generically called the {\it full counting statistics} (FCS) \cite{allez,Dhar2018,Lacroix_rotating,castillo,akemann1,akemann}. Note that FCS has also been widely studied in the related one-dimensional log-gas, that arises in the classical ensembles of random matrix theory \cite{Mehtabook,Forrester}, see e.g. \cite{MNSV2009,MNSV2011,MV2012,MMSV2014}. In fact, for the log-gas, several examples of {\it smooth} linear statistics have been considered, \blue{beyond the center of mass \cite{NM2009}}. For instance, the case $f(x) = x(1-x)$ has been studied in the context of chaotic transport through a cavity where the positions $x_i$'s, with $0 \leq x_i \leq 1$, map onto the eigenvalues of an $N \times N$ Jacobi random matrix \cite{Beenaker1963,sommers,Khor,OK2008,VMB2008,VMB2010,DMTV2011}.
The case $f(x) = x^q$ has been considered in the context of the R\'enyi entropy in a random pure state of a bipartite system -- in this case $q>0$ is the R\'enyi index~\cite{Nadal1,Nadal2}. \blue{Yet another instance is the case $f(x) = 1/x$ for Wishart-Laguerre ensemble that has been studied in the context of the Wigner time-delay distribution \cite{TM2013}.}}

In the case where the ${\bx_i}$'s are independent and identically distributed (IID) random variables, assuming that $f(\bx_i)$ has well defined first and second moments, the mean of ${\cal L}_N$ is clearly of order $O(N)$ while the typical fluctuations, which are of order $O(\sqrt{N})$, are simply Gaussian, due to the well known Central Limit Theorem (CLT). In fact, with the additional assumption that the $q$-th cumulant $\kappa_q$ of $f(\bx_i)$ exists, it is then easy to see that the $q$-th cumulant of ${\cal L}_N$ is just $N\, \kappa_q$ for IID variables. However, for the Coulomb gas, because of the long-range interactions [see Eq. (\ref{Coulomb_int}) below], the positions of the particles $\bx_i$'s are strongly correlated and these standard results for IID can not be used. Nevertheless, for smooth enough functions (which thus excludes the case of the full counting statistics where $f(\bx)$ is an indicator function), a CLT has also been established 
for several classes of external potentials and in various dimensions~$d$ \cite{Forrester_91,RiderVirag2007,Ameur2011,Leble2018,BBNY,Flack2023,Serfaty2023}. These works yield quite general formulae for the first two cumulants of ${\cal L}_N$ to leading order at large $N$, which in some cases lead to very explicit expressions.
These results are valid in the high density regime, where possible Wigner crystallization is avoided.
By contrast there does not seem to be analogous results for the higher order cumulants of the linear statistics in this dense regime.
The aim of this paper is to establish such results in any dimension $d$ in the case 
of rotationaly invariant external potential and function $f(\bx) = f(|\bf x|)$.

We now turn to the definition of the model and of the observable as well as the presentation of the main results.

\subsection{Model}

We consider a $d$-dimensional Coulomb gas of $N$ particles at positions $\bx_1,\dots,\bx_N$ in
the presence of an external potential. At equilibrium in the canonical ensemble 
at inverse temperature $\beta$ the joint probability distribution function (PDF) of the positions 
${\cal P}(\bx_1,\ldots,\bx_N)$
is given by
\begin{align}
  {\cal P}(\bx_1,\ldots,\bx_N) = \frac{1}{Z_{N,U}} e^{- \beta {\cal E}(\bx_1,\ldots,\bx_N)}
  \,,\label{eq:pdf}
\end{align}
where $\beta$ is the inverse temperature and the total energy reads
\begin{align} \label{energy}
  {\cal E}(\bx_1,\ldots,\bx_N) = \sum_{1\leq i<j\leq N} V_d(|\bx_i-\bx_j|) + N \sum_{k=1}^N U(\bx_k)\,.
\end{align}
Here $V_d(x)$ is the Coulomb interaction potential in $d$ dimensions
\begin{align}\label{Coulomb_int}
  V_d(x) = \left\{\begin{array}{ll}
    \frac{x^{2-d}}{d-2} & d\neq 2\,,\\
    -\log(x) &d=2\,,
  \end{array}\right.
\end{align}
while $U(\bx)$ is an external potential, which is assumed to be smooth and confining, so that the particles remain in the vicinity of the origin
\footnote{Here one assumes that $U(\bx) \gg \log |\bx|$ at large $|\bx|$, see e.g., \cite{saff}.}.
In most of the applications below we will consider the case of a rotationally invariant potential $U(\bx)=U(|\bx|)$. 
Note that in Eq. (\ref{energy}) the effective potential is actually $N\,U(\bx)$, such that the equilibrium average density of particles has a support of order $O(1)$ in the limit $N \to \infty$. Finally, $Z_{N,U}$ in Eq. (\ref{eq:pdf}) denotes the partition function of the system, i.e.,   
\bea \label{eq:Z}
Z_{N,U} = \int_{\mathbb{R}^d} d\bx_1 \cdots \int_{\mathbb{R}^d} d\bx_N \, e^{-\beta\left[ \sum_{1\leq i<j\leq N} V_d(|\bx_i-\bx_j|) + N \sum_{k=1}^N U(\bx_k)\right]} \;.
\eea
In the particular case of $d=2$ with $\beta=2$ and $U(\bx)=\frac{1}{2}\bx^2$, the joint PDF in Eq.~\eqref{eq:pdf} also
describes the joint PDF of the eigenvalues $\bx_i \to z_i \in \mathbb{C}$ of the complex Ginibre ensemble (in complex
number notations)
\begin{align}
  P(z_1,\ldots,z_N) = \frac{1}{Z_N}\prod_{i<j}|z_i-z_j|^2 \prod_{i=1}^N e^{-N|z_i|^2}\,.
\end{align}

An important quantity is the mean density defined as
\be 
\bar \rho_N(\bx) = \left\langle \frac{1}{N} \sum_{i=1}^N \delta(\bx-\bx_i) \right\rangle_{U}
\ee 
where $\langle \cdots \rangle_U$ denotes the average over the joint PDF in (\ref{eq:pdf}).
For later purpose it is convenient to indicate the external potential as a subscript.

\subsection{Equilibrium density in the large $N$ limit}\label{sec:density}

Let us recall how to obtain the equilibrium density in the large $N$ limit. 
It is convenient to introduce the empirical density defined as 
\be 
\rho_{N}(\bx) =  \frac{1}{N} \sum_{i=1}^N \delta(\bx-\bx_i)  \;,
\ee 
where the ${\bx}_i$'s are distributed according to Eq. (\ref{eq:pdf}). 
The energy in Eq. (\ref{energy}) can be rewritten as a functional of the
density, which in the large $N$ limit takes the form (for a pedagogical derivation
see e.g. Ref. \cite{SatyaDavid})
\bea  
&& {\cal E}(\bx_1,\ldots,\bx_N) = N^2 E[\rho_N] + O(N) \\
&& E[\rho] = \frac{1}{2} \int_{\mathbb{R}^d} d\bx \int_{\mathbb{R}^d} d\bx' \rho(\bx) \rho(\bx') V_d(|\bx-\bx'|) + \int_{\mathbb{R}^d} d\bx \, \rho(\bx)  U(\bx)\,.
\eea   
As $N \to \infty$, $\rho_{N}(\bx)$ coincides with its average $\bar \rho_{N}(\bx)$,
and both converge to the equilibrium density $\rho_{\rm eq}(x)$ which is given by the minimizer of $E[\rho]$, 
under the additional constraint $\int_{\mathbb{R}^d} \rho(\bx) d\bx = 1$. 

To determine the minimizer one takes a functional derivative of $E[\rho]$ with respect to (w.r.t.) $\rho(\bx)$. This leads to
the following condition
\bea \label{mini1}
 \int_{\cal{D}} d\bx'  \rho_{\rm eq}(\bx') V_d(|\bx-\bx'|) +  U(\bx) + \lambda = 0 \;, \quad 
\eea
which is valid for any $\bx \in {\cal D}$, where ${\cal D}$ denotes the support of $\rho_{\rm eq}$ (with $\rho_{\rm eq}(\bx) = 0$ for $\bx \notin {\cal D}$) and 
$\lambda$ is a Lagrange multiplier enforcing the normalization of $\rho_{\rm eq}(\bf x)$. Taking the Laplacian w.r.t. ${\bx}$ on both sides of Eq. (\ref{mini1}) and using that $V_d(|\bx|)$ satisfies the Poisson equation, i.e.,
\bea \label{Poisson}
\Delta V_d(|\bx|) = -\Omega_d \, \delta(\bx) \quad, \quad \Omega_d = \frac{2\pi^{d/2}}{\Gamma(d/2)} \;,
\eea
one obtains the equilibrium measure for $\bx \in {\cal D}$ as
\be \label{rho_eq}
\rho_{\rm eq}(\bx) = \frac{1}{\Omega_d} \Delta U(\bx) \;.
\ee 
Although this equation is always correct in any $d$, for a general $U(\bx)$ the hard problem 
is to determine the support ${\cal D}$, also called the ``droplet''. Note that the support is such that $\Delta U(\bx) \geq 0$ for all $\bx \in {\cal D}$. 

In the case of a rotationally invariant potential $U(\bx) = U(|\bx|)$, which we will focus on below,
the density is also rotationally
invariant and given by
\be 
\rho_{\rm eq}(\bx) = \rho_{\rm eq}(|\bx|) \quad , \quad  \rho_{\rm eq}(x) = \frac{1}{\Omega_d} \frac{1}{x^{d-1}}(x^{d-1} U'(x))' \;.
\ee 
If the potential is smooth, the droplet ${\cal D}$ is
a sphere of radius $R$. Its value is determined by the normalization condition
$\int d\bx \, \rho_{\rm eq}(\bx)= \Omega_d \int_0^R dx x^{d-1} \rho_{\rm eq}(x)  
= 1$ which leads to the condition
\be \label{norm1}
U'(R) R^{d-1} = 1 \;,
\ee
where we assumed that $U'(x) x^{d-1}$ vanishes at $x=0$. 
In the following, to ensure unicity of $R$ in \eqref{norm1}, and the validity of the saddle point methods 
used below (to be on the safe side), we will assume a slightly stronger condition, namely
\be \label{conditions}
0 < U'(0) < + \infty \quad , \quad U''(x) >0  \;,
\ee 
where the second condition needs to hold only in a region including the droplet.

\subsection{Linear statistics}

We are interested in the linear statistics, i.e. the fluctuations of ${\cal L}_N$ defined as
\begin{align}
  {\cal L}_N = \sum_{i=1}^N f(\bx_i)\,,\label{eq:lin}
\end{align}
where $f(\bx)$ is a given arbitrary sufficiently smooth function (see below for more precise conditions). In particular, the class of functions considered here does not include indicator functions which appear in the
counting statistics problem. Again, we will focus below 
on rotationally invariant linear statistics $f(\bx )=f(|\bx|)$.

To study the distribution of the linear statistics in (\ref{eq:lin}), one defines its cumulant generating function CGF $\chi(s,N)$ as
\begin{align}
 \chi(s,N) =  \log \langle e^{- N s \, {\cal L}_N } \rangle_U = 
 \log \langle e^{- N s \sum_{i=1}^N f(\bx_i)}\rangle_U \,,\label{eq:chidef}
\end{align}
where we recall that the average $\langle \cdots \rangle_U$ is taken over the joint PDF in (\ref{eq:pdf}). 
The cumulants of ${\cal L}_N$, denoted as $\langle {\cal L}_N^q \rangle_c$, are then obtained from an expansion in
the parameter $s$
\be \label{cum_gen}
\chi(s,N) = \sum_{q \geq 1} \frac{(-1)^q}{q!} N^q s^q \langle {\cal L}_N^q \rangle_c \;,
\ee 
where $s$ is sufficiently small to ensure the convergence of the series. 

\subsection{Main results}  

Let us summarize our main results. In Section \ref{sec:gen} we start with a generic potential $U(\bx)$ and function $f(\bx)$. 
We show that the calculation of the CGF $\chi(s,N)$ in (\ref{eq:chidef}) becomes equivalent to studying a Coulomb gas in a
tilted $s$-dependent potential $\tilde U_s(\bx) = U(\bx) + \frac{s}{\beta} f(\bx)$. 
From this observation, we obtain a formula for the derivative $\partial_s \chi(s,N)$ in the large $N$ limit,
given in \eqref{eq:chiss3}. It involves the droplet ${\cal D}_s$ associated with the tilted potential,
which is in principle determined from \eqref{norms}. 
Although these expressions are exact they are very difficult to solve explicitly. 

Hence, in Section \ref{sec:rot_inv}, we specify to the rotationally invariant case $U(\bx)=U(|\bx|)$ and $f(\bx)=f(|\bx|)$, for which explicit results can be obtained. 
In that case the droplet is a sphere of radius $R_s$. One obtains a simple formula for the second derivative of
the CGF in the large $N$ limit, which reads 
\begin{eqnarray} \label{derivative_G_intro}
\partial^2_s\chi(s,N) \simeq N^2 \beta^{-1} \int_0^{R_s} dx x^{d-1} [f'(x)]^2 \;,
\end{eqnarray}
where $R_s$ is determined by
\be 
  R_s^{d-1}\left( U'(R_s) + \frac{s}{\beta} f'(R_s)\right)= 1\,.\label{eq:Ruc_intro}
\ee 
This allows to obtain iteratively the cumulants $\langle {\cal L}_N^q\rangle$ in the large $N$ limit.
The second cumulant reads
\begin{eqnarray} \label{secondcum_intro}
\langle {\cal L}_N^2 \rangle_c \simeq \beta^{-1} \int_0^{R} dx \, x^{d-1} [f'(x)]^2  \quad , \quad U'(R) R^{d-1} = 1 \;.
\end{eqnarray}
This formula agrees with previous results obtained in the math literature in $d=2$ in \cite{RiderVirag2007,Ameur2011,Leble2018} (in particular, the complex Ginibre ensemble studied in \cite{RiderVirag2007,Ameur2011} corresponds to $\beta = 2$ and $U(x) = x^2/2$). It also agrees with related results obtained in $d>2$, see \cite{Serfaty2023}. One can also check that this formula yields back the result obtained in Ref.~\cite{Flack2023} in $d=1$. 

To our knowledge higher order cumulants of the linear statistics have not been studied in general dimension. 
Remarkably, although the variance of ${\cal L}_N$ depends on the value of the function $f'(x)$ on the {\it entire} droplet,  
we find from \eqref{derivative_G_intro} 
that the higher cumulants only depend on the function $f'(x)$
and its higher derivatives at $x=R$, the {\it boundary} of the droplet. Their explicit general expressions become more
and more involved with the order, and we display here the result only for the third cumulant (the fourth cumulant is
given in \ref{app_cumul})
\be 
\langle {\cal L}_N^3 \rangle_c \simeq \frac{1}{\beta^2 N} 
\frac{R^{d-1} f'\left(R\right)^3}{
   U''\left(R\right)+\frac{d-1}{R^d}}   \quad , \quad U'(R) R^{d-1} = 1 \;.
\ee 
In addition, at the end of Section \ref{sec:rot_inv}, we obtain via a Legendre transform, an expression for the large deviation
rate function of the full probability distribution of the linear statistics ${\cal L}_N$
at large $N$. The detailed results are given in \ref{app:legendre}, and are compared to known results obtained in $d=1$ \cite{Flack2023}. 

In Section \ref{sec:ginibre}, we focus on the case $d=2$ where the Coulomb gas has logarithmic interactions. We
further specify to $\beta=2$ in which case the problem 
has also a determinantal structure (this corresponds to the well known complex Ginibre ensemble of random matrix theory \cite{Mehtabook,Forrester}). 
This allows to compute the CGF using the tools of determinantal
point processes, which provides an alternative and completely different method to 
compute the CGF, and in particular its large $N$ limit. This calculation is performed
in Section \ref{sec:ginibre} and found to agree perfectly with the result of the Coulomb gas method. Finally, we conclude in Section \ref{Sec:conclusion}, while technical details have been left in Appendices.

\section{General framework} \label{sec:gen}

From the definition of the CGF \eqref{eq:chidef} and the definition of the partition function in \eqref{eq:Z}, 
we observe that one can write
\be 
\chi(s,N) = \log \frac{ Z_{N,U+ s f/\beta}}{Z_{N, U}} \;,
\ee 
where the numerator is the partition sum of the Coulomb gas in a shifted external potential
$\tilde U_s(\bx) = U(\bx) + \frac{s}{\beta} f(\bx)$. Taking a derivative w.r.t. $s$ we obtain 
\begin{align}
  \partial_s\chi(s,N) = - N \left\langle \sum_{i=1}^N f(\bx_i)\right\rangle_{\tilde U_s}\,, \label{eq:chiss}
\end{align}
where $\langle \cdots\rangle_{\tilde U_s}$ denotes the average for the Coulomb gas in the shifted external potential~$\tilde U_s$. To proceed, it is convenient to introduce the mean density in the shifted potential 
\begin{align}
    \bar\rho_{s,N}(\bx) = \left\langle \frac{1}{N} \sum_{i=1}^N \delta(\bx-\bx_i) \right\rangle_{\tilde U_s}\,\label{eq:rhob} \;.
  \end{align}
The CGF in (\ref{eq:chiss}) can then be written as
  \begin{align}
    \partial_s\chi(s,N) = - N^2\,\int_{\mathbb{R}^d}d \bx \, \bar\rho_{s,N}(\bx) f(\bx)\,.\label{eq:chiss2}
  \end{align}
 
In the large $N$ limit we know from Section \ref{sec:density} that the equilibrium density in the presence of the shifted potential converges
to 
\be \label{rho_sN}
\bar\rho_{s,N}(\bx) \underset{N \to \infty}{\longrightarrow} \rho_{{\rm eq},s}(\bx) = \frac{1}{\Omega_d} \Delta \left( U(\bx) + \frac{s}{\beta} f(\bx) \right) \;,
\ee 
which has a support which we denote ${\cal D}_s$. An additional constraint is the normalization condition
\be \label{norms}
\frac{1}{\Omega_d}  \int_{{\cal D}_s} d \bx \, \Delta \left( U(\bx) + \frac{s}{\beta}  f(\bx) \right)  = 1 \;.
\ee 
By injecting the result (\ref{rho_sN}) into \eqref{eq:chiss2}, 
the CGF can then be determined, to leading order for large $N$, from  
 \begin{align}
    \partial_s\chi(s,N) \simeq - \frac{N^2}{\Omega_d}  \int_{{\cal D}_s}d \bx \,  f(\bx) \Delta \left( U(\bx) + \frac{s}{\beta}  f(\bx) \right)   \,.\label{eq:chiss3}
  \end{align}
From Eq. (\ref{eq:chiss3}), it follows that the first cumulant [i.e., corresponding to $q=1$ in (\ref{cum_gen})], is given by $\langle {\cal L}_N \rangle = -(1/N) \partial_s \chi_s(s,N) \vert_{s=0}$, which leads to
\bea \label{av_LN}
\langle {\cal L}_N\rangle \simeq N \frac{1}{\Omega_d}  \int_{{\cal D}}d \bx \,  f(\bx) \Delta U(\bx) = N\,\int_{{\cal D}}d \bx \, \rho_{\rm eq}(\bx) f(\bf x) \;,
\eea 
where, in the last equality, we have used the expression of $\rho_{\rm eq}(\bx)$ given in Eq. \eqref{rho_eq}. The last expression in \eqref{av_LN} is of course the expected formula for the average value of the linear statistics in the large $N$ limit.

In principle these equations, together with the minimization conditions such as \eqref{mini1} extended to the tilted potential $\tilde U_s(\bx)$ allow to determine ${\mathcal D}_s$ and eventually $\chi(s,N)$. Unfortunately, it is
a very hard problem to obtain ${\cal D}_s$ without assuming any symmetry, hence this is as far as we can go for general
$f(\bx)$ and $U(\bx)$.

\section{Rotationally invariant case}\label{sec:rot_inv}

\subsection{Cumulant generating function}

We now consider the case where both $U$ and $f$ are rotationally invariant, i.e.,
\be \label{U_f_inv}
U(\bx) = U(|\bx|) \quad , \quad f(\bx) = f(|\bx|)  \;.
\ee 
In that case the support ${\cal D}_s$ is a sphere of radius $R_s$ (and in $d=1$ it is the interval $[-R_s,R_s]$).
The density is rotationally invariant and reads
\be \label{rho_eqs}
 \rho_{{\rm eq},s}(\bx)  = \rho_{{\rm eq},s}(|\bx|) \quad , \quad  \rho_{{\rm eq},s}(x)  = \frac{1}{\Omega_d} \frac{1}{x^{d-1}}\left(x^{d-1} U'(x)+ \frac{s}{\beta}  f'(x)\right)' \;.
\ee 
One can then use the relation (\ref{norm1}) with the substitution $U \to \tilde U_s = U + s\,f/\beta$ -- corresponding to the normalisation condition of $\rho_{{\rm eq},s}(\bx)$ -- to obtain the equation that determines $R_s$, namely
\be 
  R_s^{d-1}\left( U'(R_s) + \frac{s}{\beta} f'(R_s)\right)= 1\,.\label{eq:Ruc}
\ee 
In the following we assume that there is a single solution to this equation. Under the conditions 
\eqref{conditions} we expect that this is realized for a large class of smooth functions $f(x)$
at least for $s$ sufficiently close to zero (which is what is needed here to compute the cumulants). In this case, using Eq. (\ref{rho_eqs}), the relation in \eqref{eq:chiss3} reads
\be
   \partial_s\chi(s,N) \simeq - N^2 \int_0^{R_s} dx \, f(x) \, \left( x^{d-1}U'(x) + \frac{s}{\beta} x^{d-1} f'(x)\right)' \,.\label{eq:G}
\ee
Integrating by parts and using the condition \eqref{eq:Ruc}, it can be rewritten as
\be \label{eq:G2}
\partial_s\chi(s,N) \simeq N^2 \int_0^{R_s} dx \, f'(x) \, \left( x^{d-1}U'(x) + \frac{s}{\beta} x^{d-1} f'(x)\right) - N^2 f(R_s) \;,
\ee
where we have assumed that 
\be 
\lim_{x \to 0^+} \left( x^{d-1}(U'(x) + \frac{s}{\beta}  f'(x)) \right) = 0  \;. \label{eq_condition}
\ee
Hence in $d=1$ we need that $f'(0)=0$ and $U'(0)=0$. Taking another derivative
with respect to $s$, one observes that the terms proportional to $dR_s/ds$ cancel exactly
due again to the condition \eqref{eq:Ruc}. One finally finds the leading behavior in the large $N$ limit
\begin{eqnarray} \label{derivative_G}
\partial^2_s\chi(s,N) \simeq N^2 \beta^{-1} \int_0^{R_s} dx x^{d-1} [f'(x)]^2 \;.
\end{eqnarray}

This formula is the starting point for the evaluation of the cumulants $\langle {\cal L}_N^q \rangle_c$ for $q \geq 2$
in the next section. The higher cumulants $q \geq 3$ will be obtained by taking further derivatives of \eqref{derivative_G}
with respect to $s$. One sees, remarkably, that they will involve only the  ``local'' behavior of $f'(x)$ and $U'(x)$ close to the edge 
of the density at $x=R$, where we recall that $R$ is determined by \eqref{norm1} -- while $R_s$ is given by \eqref{eq:Ruc}. 

\subsection{Cumulants}

We can now give some explicit formula for the cumulants. Let us recall that they are obtained from the relation \eqref{cum_gen},
equivalently
% \be 
% \chi(s,N) = \sum_{q \geq 1} \frac{(-1)^q}{q!} N^q s^q \langle {\cal L}_N^q \rangle_c \;.
% \ee 
\be 
\langle {\cal L}_N^q \rangle_c = \frac{1}{N^q} (-1)^q \partial_s^q \chi(s,N)|_{s=0} \;. \label{cumder}  
\ee 

The first cumulant is obtained by setting $s=0$ in \eqref{eq:G}, where $R_0=R$ is determined by \eqref{norm1}. It reads
\be \label{firstcum}
   \langle {\cal L}_N \rangle \simeq N \int_0^{R} dx \, f(x) \, \left( x^{d-1}U'(x) \right)'  \quad , \quad U'(R) R^{d-1} = 1
\ee
The second cumulant is obtained by setting $s=0$ in \eqref{derivative_G} and reads
\begin{eqnarray} \label{secondcum}
\langle {\cal L}_N^2 \rangle_c \simeq \beta^{-1} \int_0^{R} dx x^{d-1} [f'(x)]^2  \quad , \quad U'(R) R^{d-1} = 1 \;.
\end{eqnarray}
%This formula agrees with previous results obtained in the math literature, both in $d=2$ in \cite{RiderVirag2007,Ameur2011,Leble2018} (in particular, the complex Ginibre ensemble studied in \cite{RiderVirag2007,Ameur2011} corresponds to $\beta = 2$ and $U(x) = x^2/2$) and for $d>2$ in \cite{Leble2018,Armstrong2021}. One can also check that this formula yields back the result obtained in Ref. \cite{Flack2023} in $d=1$. 
\\

Using \eqref{derivative_G} and \eqref{cumder}, the  higher cumulants for $q \geq 3$ are obtained from
\be \label{cumulant_expl}  
\langle {\cal L}_N^q \rangle_c \simeq \frac{(-1)^q}{\beta N^{q-2}} \, \partial_s^{q-3} \left( \frac{dR_s}{ds} R_s^{d-1} f'(R_s)^2 
 \right)\Big|_{s=0}  \;.
\ee 
To obtain explicit formula we need to compute the derivative $\frac{dR_s}{ds}$.
Taking a derivative of \eqref{eq:Ruc} with respect to $s$, and using again \eqref{eq:Ruc} to
simplify the expression one obtains
% \be \label{ders} 
% \frac{dr_s}{ds} = -\frac{T f'(r_s)}{s T
%    f''(r_s)+V''(r_s) +\beta  T/r_s^2}
% \ee 
%  \be 
% \frac{dR_s}{ds} =-\frac{R_s^d f'\left(R_s\right)}{s R_s^d
%    f''\left(R_s\right)+\beta  \left(R_s^d
%    U''\left(R_s\right)+d-1\right)}
%    \ee 
    \be  \label{dRsds} 
\frac{dR_s}{ds} =-\frac{f'\left(R_s\right)}{s 
   f''\left(R_s\right)+\beta  \left(
   U''\left(R_s\right)+\frac{d-1}{R_s^d}\right)}  \;.
   \ee 
Substituting $\frac{dR_s}{ds} $ in \eqref{cumulant_expl} we obtain the third cumulant as
\be 
\langle {\cal L}_N^3 \rangle_c \simeq \frac{1}{\beta^2 N} 
\frac{R^{d-1} f'\left(R\right)^3}{
   U''\left(R\right)+\frac{d-1}{R^d}}   \quad , \quad U'(R) R^{d-1} = 1 \label{third_cumul} \;.
\ee 
It is possible to compute systematically the higher cumulants as functions of the derivatives of $U(x)$ and $f(x)$ at
$x=R$, although the
expressions become more and more bulky. This is done in \ref{app_cumul} where the
explicit expression of the fourth cumulant is displayed. 

\subsection{Full probability distribution of the linear statistics}

Our results also allow to obtain information about the PDF ${\cal P}({\cal L}_N)$
of the linear statistics
${\cal L}_N = \sum_i f(|\bx_i|)$. Indeed at large $N$ we expect that it takes the large deviation form
\be \label{PP}
{\cal P}({\cal L}_N) \sim e^{- N^2 \Psi(\Lambda) } \quad , \quad \Lambda=\frac{1}{N} {\cal L}_N = \frac{1}{N} \sum_i f(|\bx_i|) \;.
\ee 
\blue{By inserting this large deviation form (\ref{PP}) in the definition of the CDF in Eq. (\ref{eq:chidef}) and performing the change of variables ${\cal L}_N \to \Lambda$ one obtains
\bea \label{eq:saddle_point}
\chi(s,N) \approx \log \left( \int_0^\infty d\Lambda \; e^{-N^2 \left( \Psi(\Lambda) + s\, \Lambda\right)}\right) \;.
\eea 
For large $N$ the integral over $\Lambda$ can be evaluated by the saddle point method, leading to the relation 
%The rate function $\Psi(\Lambda)$ is related to the CGF computed by Legendre inversion of
\be 
\lim_{N\to \infty}\, \frac{1}{N^2}\chi(s,N) = - \min_{\Lambda \in \mathbb{R^+}} ( \Psi(\Lambda) + s \,\Lambda) \;.  \label{legendre00} 
\ee 
One can thus extract $\Psi(\Lambda)$ from our result for $\chi(s,N)$ by Legendre inversion of (\ref{legendre00}).} This is performed in the \ref{app:legendre}.
The general formula is a bit complicated, and for illustration we also work out in the \ref{app:legendre} a few examples
where simple explicit formulae can be obtained.

\section{Determinantal case: $\beta=2$ in $d=2$ and the Ginibre ensemble}\label{sec:ginibre}

\subsection{Exact formula for the CGF} 

In the special case $\beta=2$ in space dimension $d=2$, with full rotational invariance,
the calculation can be performed in a completely different way using determinantal formula 
which, remarkably, gives the same result. This can be achieved for any potential $U(x)$. 
Furthermore, in the case $U(x)=\frac{x^2}{2}$ the probability weight \eqref{eq:pdf} of the particle positions in the Coulomb gas
can also be interpreted as the 
joint distribution of the eigenvalues of a complex non-Hermitian Gaussian 
random matrix in the Ginibre ensemble~\cite{Mehtabook,Forrester}. More precisely, let $(z_1,\ldots,z_N)$ denote the complex eigenvalues of the Ginibre ensemble.
Their joint PDF $p(z_1,\ldots,z_N)$, is given by
\begin{align} \label{ginibre}
  p(z_1,\ldots,z_N) = \frac{1}{Z_N}\prod_{i<j}|z_i-z_j|^2 \prod_{i=1}^N e^{-N|z_i|^2}\,,
\end{align}
where $Z_N$ is the normalisation coefficient. For simplicity we will focus here on that case, the
general $U(x)$ being discussed below and in \ref{app_det}. 

As is well known, the joint PDF \eqref{ginibre} defines a determinantal point processes \cite{Mehtabook,Forrester}. 
Using standard manipulations, exploiting the rotational symmetry of the potential and of the
linear statistics function $f(z_i) = f(|z_i|)$, one can show the following formula for the CGF in Eq. \eqref{eq:chidef}
valid for any $N$ (see e.g. \cite{Rider})
\begin{align}
  \chi(s,N) &= \sum_{\ell=0}^{N-1} \ln\left(2 N^{1+\ell}  \int_0^\infty dr \frac{r^{2 \ell+1}}{\ell!}  e^{-N r^2 - s N f(r)}\right)\,.\label{eq:chis}
\end{align}
It is a simple consequence of the determinantal structure and of the Cauchy-Binet formula -- we recall it in \ref{app_det}
(and extend to more general $U(x)$). 
For the case $U(x)=x^2/2$ discussed here, it can also be derived from quantum mechanics of fermions in a rotating trap \cite{Lacroix_rotating,Smith_rotating,Manas_rotating1,Manas_rotating2}.
In that framework the label $\ell$ has the interpretation 
of angular momentum and the integrand identifies with the radial component of the eigenstates of
angular momentum $\ell$ 
within the lowest Landau level
(see \cite{Mehtabook,Lacroix_rotating,Smith_rotating,Manas_rotating1,Manas_rotating2} for more details).

% \begin{align}
%   \chi(s,N) &= \sum_{\ell=1}^N \ln\left(N^\ell \int_0^\infty du \frac{u^{\ell-1}}{\Gamma(\ell)}  e^{-Nu -\frac{s}{N}f(\sqrt{u})}\right)\,.\label{eq:chis}
% \end{align}
% \newpage

\subsection{Asymptotic analysis for large $N$} \label{4.2}

% where $\mathcal{F}$ is the primitive of $\mathcal{G}$ in (\ref{eq:scalG}).

Let us now analyse \eqref{eq:chis} in the large $N$ limit. It is easy to see that for large $N$ the discrete sum is dominated 
by terms with $\ell = O(N)$. We thus perform the change of variable $\ell=N \lambda$ and 
replace the sum in (\ref{eq:chis}) by an integral. This leads to the estimate
\be \label{00}
\chi(s,N) \simeq N \int_0^1 d\lambda  \ln\left(2 N^{N \lambda+1}  \int_0^\infty dr \frac{r^{2 N \lambda+1}}{(N \lambda)!}  e^{-N r^2 - s N f(r)}\right) \;.
\ee 
We can now use that for large $N$
\bea 
\frac{2 N^{\lambda N + 1}}{(\lambda N)!}  \simeq \frac{\sqrt{2N}}{\sqrt{\pi \lambda}}
e^{N( \lambda - \lambda \log \lambda)} \;.
\eea 
This leads to
% \be 
% \chi(s,N) \simeq N \int_0^1 d\lambda  \ln\left( \frac{\sqrt{2N}}{\sqrt{\pi \lambda}}
%   \int_0^\infty dr r 
%   e^{-N (r^2 + s  f(r) - 2 \lambda \log r - \lambda + \lambda \log \lambda)}
%   \right) 
% \ee 
\bea 
&& \chi(s,N) \simeq N \int_0^1 d\lambda  \ln\left( \frac{\sqrt{2N}}{\sqrt{\pi \lambda}}
  \int_0^\infty dr \,r\; 
  e^{-N \phi_{s,\lambda}(r) }
  \right)  \\
&& {\rm where}\;\;
 \phi_{s,\lambda}(r)  = r^2 + s  f(r) - 2 \lambda \log r - \lambda + \lambda \log \lambda \label{phi} \;.
\eea   
The integral over $r$ can be performed using the saddle point method for any fixed $\lambda$ and~$s$.
One can check that for $s=0$ the saddle point is at $r=\sqrt{\lambda}$. Performing the Gaussian integral around the saddle point 
one finds that the leading term in $\chi(s,N)$ vanishes, 
as it should from normalization. \blue{For general $s$ the saddle point method shows that at large $N$ 
\be
\chi(s,N)  \simeq N^2 \mathcal{F}(s) + o(N^2) 
\ee 
with 
\be
   \mathcal{F}(s)  = - \int_0^1 d\lambda\,\min_{r\geq 0}[\phi_{s,\lambda}(r)] = - \int_0^1 d\lambda \, \phi_{s,\lambda}(r_{s,\lambda}) \,,\label{eq:F}
\ee 
where $r_{s,\lambda}>0$ minimises $\phi_{s,\lambda}(r)$, i.e., it is the solution of
\begin{eqnarray}
\partial_r \phi_{s,\lambda}(r)|_{r=r_{s,\lambda}} = 0 \quad \Longleftrightarrow \quad 
r_{s,\lambda}^2 + \frac{s}{2} r_{s,\lambda} f'(r_{s,\lambda}) = \lambda \;. \label{saddle}
\end{eqnarray}
Taking one derivative with respect to $s$ in Eq. (\ref{eq:F}), using $\partial_r \phi_{s,\lambda}(r)|_{r=r_{s,\lambda}}=0$ together with 
the explicit dependence of $\phi_{s,\lambda}(r)$ on $s$ in~\eqref{phi}, one finds
\begin{eqnarray}
\partial_s  \mathcal{F}(s) = - \int_0^1 d\lambda f(r_{s,\lambda}) \label{48} \;.
\end{eqnarray}
}
%where $r_{s,\lambda}>0$ defined in (\ref{saddle}).
%\begin{eqnarray}
%\partial_r \phi_{s,\lambda}(r)|_{r=r_{s,\lambda}} = 0 \quad \Longleftrightarrow \quad 
%r_{s,\lambda}^2 + \frac{s}{2} r_{s,\lambda} f'(r_{s,\lambda}) = \lambda \;. \label{saddle}
%\end{eqnarray}
%
%
Hence, for any fixed $s$, performing in Eq. \eqref{48} the change of variable $\lambda \to r_{s,\lambda}$, 
and expressing $d\lambda$ using \eqref{saddle} one obtains
\begin{eqnarray} \label{eqFu}
\partial_s  \mathcal{F}(s) = - \int_0^{R_s} d\left( r_{s,\lambda}^2 + \frac{s}{2} r_{s,\lambda} f'(r_{s,\lambda}) \right) f(r_{s,\lambda}) \;,
\end{eqnarray}
where the upper bound of the integral $r_{s,\lambda=1}$ exactly identifies with the quantity $R_s$ defined in the previous section, see Eq. \eqref{eq:Ruc}.
Indeed, one can compare the condition~\eqref{saddle} with Eq. \eqref{eq:Ruc}, which becomes in $d=2$, for $U(x)=x^2/2$ and for $\beta=2$
\be 
R_s^2 + \frac{s}{2} R_s f'(R_s) = 1  \;. \label{Rs2} 
\ee 
In conclusion, the result of this method can thus be written as
\be
\frac{1}{N^2} \partial_s \chi(s,N) \simeq \mathcal{F}(s) = -\int_0^{R_s} dx \partial_x\left[x^2 + \frac{s}{2} x\,f'(x)\right] f(x)\,,\label{eq:G2}
\ee
where $R_s$ is the solution of \eqref{Rs2}. This precisely coincides with the result obtained with the method presented in the previous sections \ref{sec:gen} and \ref{sec:rot_inv}, i.e., Eq.
\eqref{eq:G} where one sets $U(x)=x^2/2$ and $\beta=2$. 
\\

{\bf General potential $U(x)$}: In \ref{app_det} the formula \eqref{eq:chis} is extended to the case of
arbitrary $U(x)$. The generalisation of the computations performed above is then immediate, i.e., \eqref{eq:F} still holds, with now the saddle point function
\be 
 \phi_{s,\lambda}(r)  = 2 U(r) + s  f(r) - 2 \lambda \log r - b(\lambda) \label{phi2}
\ee 
where $b(\lambda)=\min_{r>0} ( 2 U(r) - 2 \lambda \log r)$. 
The same manipulations lead again to \eqref{eq:G}, with $\beta=2$, $d=2$ and a general $U(x)$.
Note that in practice we need to ensure that there is a unique minimum to $\phi_{s,\lambda}(r)$
as a function of $r$,
for any $\lambda \in ]0,1[$ and $s$ in the vicinity of $s=0$. This should be ensured
by the conditions \eqref{conditions}, and for smooth functions $f(r)$. 
\\

{\bf Remark about the case symplectic Ginibre ensemble (GinSE):} Interestingly, it turns out that the same computation can be done for the GinSE, for which the joint PDF of the eigenvalues (given e.g., in Eq. (2.2) in \cite{akemann}) is not exactly of the Coulomb gas form as in Eq. (\ref{eq:pdf})
\footnote{It can be seen as a Coulomb gas with an additional image interactions between the charges.}. Nevertheless, the CGF can be obtained for any finite $N$ and $f(r)$ as \cite{Rider}
\bea \label{GSE}
\chi(s,N) = \log  \left( 
\prod_{j=1}^N \frac{ \int_0^{+\infty} dr r^{4j-1} \, e^{-2 N r^2  - N s f(r)} }{\int_0^{+\infty} dr r^{4j-1}  e^{-2 N u^2} } \right) \;.
\eea
The same saddle point method as for $\beta=2$ can thus be used, leading to a similar formula for the cumulants. 
It has the same salient feature that all cumulants of order $q \geq 3$ depend only on the values
of $f'(x)$ and its derivatives at the boundary of the droplet.

\section{Conclusion}\label{Sec:conclusion}

In this paper we have studied the $d$-dimensional 
Coulomb gas in an external potential at equilibrium at inverse temperature $\beta$.
We have considered the large $N$ limit where the support
of the equilibrium density is a single droplet. In that limit, 
we have computed the higher cumulants for smooth linear statistics.
In the rotationally invariant case where the droplet
is a $d$-dimensional sphere, we have obtained explicit expressions for the cumulants
for arbitrary potential $U(x)$ and linear statistics function $f(x)$. The remarkable 
property unveiled here is that the cumulants of order $q \geq 3$ depend only on
the function $f'(x)$ and its derivatives on the boundary of the droplet. 
In the case of $d=2$, $\beta =2$ and $U(x)=x^2/2$, the system is
determinantal and corresponds to the complex Ginibre ensemble of random matrices.
Using methods of determinantal point processes we obtained the same
formula for the cumulants from a completely different approach.

The present work could be extended in several directions. 
An interesting problem is the case where $f(\bx)$ is not invariant under rotation, even for a rotationally invariant
potential $U(\bx)=U(|\bx|)$. In particular for $d=2$, $\beta = 2$ and $U(x)=x^2/2$ (i.e., for the complex Ginibre ensemble) there is an explicit formula \cite{Forrester_91,RiderVirag2007,Ameur2011} for the second cumulant at large $N$. 
Furthermore, in $d=2$ there are also extensions when $U(\bx)$ is itself not invariant under rotation, which
involve conformal maps \cite{Ameur2011,Leble2018} (see also extension for $d>2$ \cite{Serfaty2023}).
It would be interesting to recover these results from our approach, and to
extend them to compute the higher cumulants, which are not known at present.
One promising direction is to study the special examples of non-rotationally invariant (e.g. elliptic) potentials 
\cite{forrester_el,oblak,difrancesco,byun}
where explicit formula exist for the droplet,
and where this program may be carried out. 

It is important to stress that the results of this paper are valid for linear statistics functions $f(x)$ which are smooth.
In the case of the counting statistics, i.e., when the function is an indicator function, hence non-smooth,
there is an intermediate regime~\cite{castillo} between the Gaussian typical behavior and the large deviation regime which was computed in \cite{allez}. 
It would also be interesting to explore the case of more general non-smooth function $f(x)$,
such as done in $d=1$ in \cite{Flack2023}. In addition, more complicated geometries, i.e. the case of several droplets 
could be interesting to investigate. 

Finally, it remains a challenge to extend these results to more general interactions, beyond the Coulomb potential, 
such as the Riesz gas \cite{lewin,riesz} or the Yukawa (i.e., screened Coulomb) potential \cite{Yukawa}.

\bigskip

{\bf Acknowledgments.} We thank P. Bourgade, B. Estienne and J.~M. St\'ephan for useful discussions. 

\appendix

\section{Higher cumulants}\label{app_cumul}

To compute higher cumulants for $q \geq 4$ systematically, it is convenient to write
\be 
\frac{dR_s}{ds} = - \frac{1}{\beta} A(R_s) 
\ee 
where the function $A(r)$ is defined as 
% \be 
% A(r) = 
% \frac{r^d f'\left(r\right){}^2}{\beta 
%    \left(f''\left(r\right) \left(r-r^d
%    U'\left(r\right)\right)+f'\left(r\right)
%    \left(r^d U''\left(r\right)+d-1\right)\right)}
%    \ee 
   \be 
A(r) = 
\frac{ f'\left(r\right){}^2}{
   f''\left(r\right) \left(r^{1-d} -
   U'\left(r\right)\right)+f'\left(r\right)
   \left( U''\left(r\right)+\frac{d-1}{r^d} \right)} \;.
   \ee 
This was obtained by eliminating the explicit $s$-dependence in the expression \eqref{dRsds}
using~\eqref{eq:Ruc}. From the change of variable from $s$ to $R_s$ we
obtain that $\partial_s = A(R_s) \partial_{R_s}$, leading to the
formula
\be 
\langle {\cal L}_N^q \rangle_c \simeq \frac{1}{\beta^{q-1} N^{q-2}} \, (A(r) \partial_r)^{q-3}  \left( A(r) r^{d-1} f'(r)^2  \right)|_{r=R} 
\quad , \quad U'(R) R^{d-1} = 1 \;.
\ee 

For the fourth cumulant one finds
\bea  
 \langle {\cal L}_N^4 \rangle_c & \simeq & \frac{1}{\beta^{3} N^{2}} \bigg( 
\frac{ \left((d-1) d-R^{d+1} U^{(3)}(R)\right) f'(R)^4 } 
{R^2 \left(U''(R)+\frac{d-1}{R^d}\right)^3}
  \\
  & + &
 \frac{  R^{d-2} f'(R)^3 \left((d-1) f'(R)+4 R f''(R)\right) }{\left(U''(R)+\frac{d-1}{R^d}\right)^2} \bigg) \nonumber
\eea  
In $d=1$ it simplifies into
\be 
\langle {\cal L}_N^4 \rangle_c  \simeq  \frac{1}{\beta^{3} N^{2}} 
\frac{f'(R)^3 \left(4 f''(R) U''(R)-U^{(3)}(R)
   f'(R)\right)}{U''(R)^3} \blue{\quad {\rm with} \quad U'(R) = 1} \;.
\ee 
In the case of the harmonic potential $U(x)= \frac{\mu}{2} x^2$ one obtains
\be 
\langle {\cal L}_N^4 \rangle_c  \simeq  \frac{1}{\beta^{3} N^{2}}  \frac{2 R^{3 d-2} f'(R)^3 \left((d-1) f'(R)+2 R
   f''(R)\right)}{d^2}
\ee 
with $\mu R^d=1$, 
which further simplifies for $d=1$ into
\be 
\langle {\cal L}_N^4 \rangle_c  \simeq  \frac{1}{\beta^{3} N^{2}}  4 R^2 f'(R)^3 f''(R) \blue{\quad {\rm with} \quad R = 1/\mu}\;.
\ee 

\vspace*{0.5cm}
\noindent{\it Special choices of $f(x)$}. For $f(x)=x$ and $d>1$ one has that
$\partial_s^2 \chi(s,N)$ is determined by eliminating $R_s$ in the system given by \eqref{eq:Ruc} and \eqref{derivative_G}, namely
\be 
\partial_s^2 \chi(s,N) = N^2 \beta^{-1} \frac{R_s^d}{d} \quad , \quad \mu R_s^d + \frac{s}{\beta} R_s^{d-1} = 1 \;.
\ee 
For $d=2$ one obtains more explicitly
\be
\partial_s^2 \chi(s,N) =  \frac{N^2 \beta^{-1}}{2 R^{-2} + s (s + \sqrt{s^2 + 4 R^{-2}})} \;.
\ee 
For general $d$, one can show that the cumulants in that case read, at leading order for large $N$, 
\bea  
\langle {\cal L}_N^q \rangle_c  & \simeq & \frac{1}{\beta^{q-1} N^{q-2} \, d^{q-2} } \prod_{j=2}^{q-2} (j d + 2-q) R^{(q-1) d + 2 - q} \\
& = & \frac{1}{\beta^{q-1} N^{q-2}  } \frac{ \Gamma\left(q-1+ \frac{2-q}{d} \right)   }{d \,\Gamma\left(2 + \frac{2-q}{d} \right) } R^{(q-1) d + 2 - q} \;.
\eea

\section{Legendre transform and the PDF of ${\cal L}_N$} \label{app:legendre}

At large $N$ the PDF of ${\cal L}_N$, namely ${\cal P}({\cal L}_N)$, takes the large deviation form given in~\eqref{PP}.
Here we obtain the rate function~$\Psi(\Lambda)$ from the CGF by Legendre inversion.
In addition we discuss a few explicit examples.

Let us recall the relation between the CGF $\chi(s,N)$ and the large deviation function $\Psi(\Lambda)$. It reads
\be 
\frac{1}{N^2}\chi(s,N) = - \min_{\Lambda \in \mathbb{R}} ( \Psi(\Lambda) + s \,\Lambda) \;.  \label{legendre0} 
\ee 
We now extract $\Psi(\Lambda)$ from our result for $\chi(s,N)$ in a parametric form.
Assuming unicity of the minimum in \eqref{legendre0} one has
\be 
\Psi'(\Lambda) = - s \;.
\ee 
One the other hand we have, taking a derivative of \eqref{legendre0} w.r.t. $s$
\be
\frac{1}{N^2} \partial_s \chi(s,N) = - \Lambda \;.
\ee 
Now Eq. \eqref{eq:G} gives us a formula for $\frac{1}{N^2} \partial_s \chi(s,N)$ as a function of $s$ and $R_s$
which we recall here
\be
  \frac{1}{N^2} \partial_s\chi(s,N) \simeq - \int_0^{R_s} dx \, f(x) \, \left( x^{d-1}U'(x) + \frac{s}{\beta} x^{d-1} f'(x)\right)' \,.\label{eq:G2}
\ee

There are several equivalent ways to eliminate some variables in order to obtain a parametric representation
of $\Psi'(\Lambda)$ or $\Psi(\Lambda)$. Let us give one here by expressing $s$ as a function of $R_s$
using \eqref{eq:Ruc}, which leads to the system 
\bea 
 \hspace*{-2.2cm} \Psi'(\Lambda) &=& - \frac{ \beta}{f'(R_s) }( R_s^{1-d} - U'(R_s))  \\
\hspace*{-2.2cm} \Lambda &=& 
\int_0^{R_s} dx \, f(x) \, (x^{d-1}U'(x) )' + \frac{1}{f'(R_s) }( R_s^{1-d} - U'(R_s))
\int_0^{R_s} dx \, f(x) \, (x^{d-1} f'(x))' \;. \nonumber 
\eea 
The relation between $\Psi'(\Lambda)$ and $\Lambda$ is then obtained parametrically by varying $R_s$. 
Once $\Psi'(\Lambda)$ is known one obtains $\Psi(\Lambda)$ as
\be
\Psi(\Lambda) = \int_{\bar{\Lambda}}^\Lambda d\lambda \, \Psi'(\lambda) ~,~ \bar \Lambda = \lim_{N \to +\infty} \frac{1}{N} \langle {\cal L}_N \rangle= 
\int_0^R dx f(x) (x^{d-1} U'(x))' ~,~ R^{d-1} U'(R) = 1 \;.
\ee 

Let us consider the harmonic potential, $U(x)=\frac{\mu}{2} x^2$. For the simple example $f(x)=x^2$, 
we obtain
\be 
\Psi'(\Lambda)  = \frac{\beta}{2} (\mu- R^{-d}) \quad , \quad \Lambda = \frac{d R^2}{d+2} \;,
\ee 
which gives
\be 
\Psi'(\Lambda)  = \frac{\beta}{2} \left(\mu- (\frac{d+2}{d} \Lambda)^{-d/2}\right) \;.
\ee 
Using $\bar \Lambda = \frac{d}{d+2} \mu^{-2/d}$ we obtain, for $d \neq 2$
\be 
\Psi(\Lambda) = \frac{\beta}{2} \left(  \frac{d^2 \mu^{1- \frac{2}{d}}}{4-d^2} +
\Lambda (\mu + \frac{2}{2-d} \left(\frac{d+2}{d} \Lambda)^{-d/2} \right) \right) \quad , \quad d \neq 2 \;,
\ee 
which for $d=1$ agrees with the result of \cite{Flack2023} (using $\mu=1/(2 \alpha)$ and $\beta=2 \alpha$) \footnote{Note that in Ref. \cite{Flack2023}, $\Lambda$ is denoted by $s$ while our $s$ is denoted is denoted by $\lambda$.}.
In $d=2$ one obtains
\be 
\Psi(\Lambda) = \frac{\beta}{4} \left(  2 \Lambda \mu - \log( 2 \Lambda \mu) - 1 \right)  \quad , \quad d = 2 \;,
\ee 
with $\bar \Lambda = \frac{1}{2 \mu}$. Note that for $d \geq 2$, $\Psi(\Lambda)$ diverges at $\Lambda=0$.

Other simple cases $f(x)=x^q$ can be worked out similarly. For $q=1$ the same method holds for any $d>1$. In $d=1$ 
however the
condition $f'(0)=0$ is not satisfied for $q=1$ [see Eq. \eqref{eq_condition}], hence our result cannot be applied in that case. 
This is because the linear statistics involves a non-analytic function ${\cal L}=\sum_i |x_i|$.
However, if one 
naively extends our result to $d=1$ one obtains
\be \label{naive} 
\Psi(\Lambda) \equiv \frac{\beta}{6 \mu} (4 \sqrt{2} (\mu \Lambda)^{3/2}- 6 \mu \Lambda + 1) \;.
\ee 
In \cite{Flack2023}
a more general calculation, allowing for a singular equilibrium
density in $d=1$, was performed for $f(x)=|x|$. It was found that the CGF $\frac{1}{N^2} \chi(s,N)$ at $N=+\infty$
exhibits a non-analyticity at $s=0$, with a delta function peak appearing
in the equilibrium density $\rho_{{\rm eq},s}(x)$ in that case, leading to a third order phase transition. 
One can check (using the correspondence $\mu=1/(2 \alpha)$ and $\beta=2 \alpha$) that \eqref{naive} agrees with that result {\it but only on the side $\Lambda< \bar \Lambda$}.
Hence our method generally misses the third order transitions which may occur when the linear
statistics function is not smooth. 

% Next we consider $f(x)=x$ (for $d>1$) 
% We obtain
% \be 
% \Psi'(\Lambda)  = \beta R (\mu- R^{-d}) \quad , \quad \Lambda = 
% \frac{R \left(d^2+\mu  R^d-1\right)}{d (d+1)}
% \ee 

\section{Determinantal case}\label{app_det}

Consider the case $d=2$ and $\beta=2$ with full rotational invariance.
Let us start from the Laplace transform of the linear statistics in (\ref{eq:chidef}),
written using complex notation for the coordinates $\bx \equiv z$
\begin{align}
  \langle e^{- N s {\cal L}_N}\rangle = \frac{1}{Z_{N,U}} \int d^2 z_1\ldots d^2 z_N \prod_{j<k}|z_j-z_k|^2 
  \prod_{j=1}^N e^{-2 N U(|z_j|) - N s f(|z_j|)}\,.
\end{align}
Using the Vandermonde form
\begin{align}
  \prod_{j<k}|z_j-z_k|^2 = \det_{1\leq j,k\leq N}z_k^{j-1} \det_{1\leq j,k\leq N}\bar z_k^{j-1} \,,
\end{align}
together with the Cauchy-Binet formula
\begin{align}
  \int d^2z_1\ldots d^2z_N \det_{1\leq j,k\leq N} f_j(z_k) \det_{1\leq j,k\leq N} g_j(z_k) \prod_{j=1}^N w(z_j) = N! \det_{1\leq j,k\leq N} \int dz f_j(z) g_k(z)w(z)\,
\end{align}
we obtain
\begin{align}
\langle e^{- N s {\cal L}_N}\rangle = \frac{N!}{Z_{N,U}} \det_{1\leq j,k\leq N} \int d^2z z^{j-1} \bar z^{k-1} e^{-2 N U(|z|) - s N f(|z|)}\,.
\end{align}
Going to polar coordinates, the integral reads
\begin{align}
  \langle e^{- N s {\cal L}_N}\rangle = \frac{N!}{Z_{N,U}}  
  \det_{1\leq j,k\leq N} \int_0^{+\infty} dr r\int_{0}^{2\pi} d\theta\, r^{j-1}  r^{k-1} e^{i\theta(j-k)} e^{-2 N U(r) - N s f(r)}\,.
\end{align}
Performing the integral over $\theta$ gives that the matrix is diagonal and therefore
\begin{align}
\langle e^{- N s {\cal L}_N}\rangle = \frac{N!}{Z_{N,U}}  
\prod_{j=1}^N \int_0^{+\infty} dr r^{2j-1}  e^{-2 N U(r)  - N s f(r)}\,.
\end{align}
By normalisation (it must equal unity for $s=0$) we obtain
\begin{align}
\langle e^{- N s {\cal L}_N}\rangle = 
\prod_{j=1}^N \frac{ \int_0^{+\infty} dr \, r^{2j-1} \,  e^{-2 N U(r)  - N s f(r)} }{\int dr r^{2j-1}  e^{-2 N U(r)} } \;.
\end{align}
The cumulant generating function (\ref{eq:chidef}) is therefore given in $d=2$, for $\beta=2$, 
for arbitrary $U(r)$, $f(r)$ and $N$ as
\begin{align}
\chi(s,N) = \sum_{j=1}^N \log  \left( \frac{ \int_0^{+\infty} dr \, r^{2j-1}  e^{-2 N U(r)  - N s f(r)} }{\int_0^{+\infty} dr r^{2j-1}  e^{-2 N U(r)} } \right) \;,
\end{align}
which is the generalization of Eq. (\ref{eq:chis}) to an arbitrary potential $U(r)$. Performing the same analysis as in Section \ref{4.2} leads to Eq. (\ref{phi2}) in the text.

\end{document}